\documentstyle[aps,prl,subfigure,twocolumn,epsfig,amsmath,amssymb]{revtex}

\begin{document}
\draft
\title{Shell structure in the  density profiles for noninteracting
fermions in anisotropic harmonic confinement}
\author{Patrizia Vignolo and Anna Minguzzi}
\address{NEST-INFM and Classe di Scienze,
Scuola Normale Superiore,
Piazza dei Cavalieri 7, I-56126 Pisa, Italy}
\maketitle
\begin{abstract}
We develop a Green's function method to evaluate the exact equilibrium
particle-density profiles of noninteracting  Fermi gases in external
harmonic confinement in any spatial dimension and for arbitrary trap
anisotropy. While in a spherically  symmetric configuration the shell
effects are negligible in the case of large number of particles, 
we find that for
very anisotropic traps the quantum effects due to single-level
occupancy and the deviations from the Thomas-Fermi
approximation are visible also for mesoscopic clouds.
\end{abstract}

Ultra-cold Fermi gases of alkali atoms are novel quantum
systems which are becoming available by the techniques of atom
trapping and cooling. The quantum degeneracy regime has been
reached with various atomic species \cite{fermi_exp} and a great
effort is now 
devoted to study the collective properties \cite{jin_exp}
and to the possibility of
observing  spatial phase separation \cite{phase_sep}
and the transition to a superfluid phase
\cite{superfluids}. Atomic Fermi gases are produced in  magnetic or
optical traps, providing an inhomogeneous external confinement
which can be well approximated as harmonic. If a single Zeeman
sublevel is populated, interparticle interactions can be neglected at
the very low temperatures of the experiments since the Pauli
principle forbids collisions in the $s$-wave channel and dipole-dipole
interactions are extremely weak for alkali-metal atoms.

The high purity of the samples and the high resolution of the
detection techniques make these systems ideal candidates for the study
of single-level quantum properties on a mesoscopic scale, such as the
shell structure in the equilibrium density profiles. In the experiments
the strength of the external confinement can be tuned to vary the
anisotropy of the harmonic trap in order to reach quasi-onedimensional and
quasi-twodimensional  configurations where, as we will show below,
the quantum effects on the equilibrium
profiles are greatly enhanced.

The exact equilibrium profiles for the particle and kinetic-energy
density of non-interacting fermions under isotropic harmonic
confinement at zero temperature have been obtained by Brack and van
Zyl \cite{brack} in $D$ dimensions ($D$=1,2,3) in terms of sums of
Laguerre polynomials. Schneider and Wallis \cite{schneider} have
computed the three-dimensional density profiles of a cigar-shaped Fermi
cloud using Hermite polynomials.

An alternative method which does not
rely on the evaluation of high-order polynomials has been 
proposed for the calculation of 1D density profiles \cite{noi}.
This Green's function method replaces the use of single-particle
orbitals in favour of matrix elements of simple operators. The basic
idea of the method is to exploit a formal analogy between the
expression of the density of single-particle states in the energy
domain and the 1D particle-density profiles in order to
rewrite the latter in terms of a suitably defined Green's function
operator in coordinate space. As originally formulated the
method is intrinsically onedimensional, since it maps the energy
variable into a 1D spatial coordinate.

In this Letter we extend the Green's function method to any spatial
dimension by making use of a systematic reduction of dimensionality in
the evaluation of the density profiles. This method allows us to treat
fully anisotropic Fermi clouds with a large number of particles and
to investigate how the shell effects are enhanced with respect to the
isotropic case. We find that very anisotropic clouds show the 
shell structure best in the tight direction 
of the confinement and that in this case the
Thomas-Fermi approximation  fails to describe the density profiles
even at large numbers of particles.

\paragraph*{Green's function method in dimension $D>1$.}

The particle density profile for noninteracting fermions under external
confinement at zero temperature is given by the zeroth moment of the
Dirac density matrix \cite{density} in terms of single-particle orbitals
$\psi_{\{i\}}(\mathbf r)$:
\begin{equation}
n_{E_F}({\mathbf r})=\sum_{\{i\}}|\psi_{\{i\}}(\mathbf r)|^2\;.
\end{equation}
Here $\{i\}$ is a complete set of quantum numbers which univocally
identify the single-particle energy levels ${\cal E}_{\{i\}}$ and
the sum runs up to the highest occupied level corresponding 
to the Fermi energy $E_F$. 

Let us first consider for simplicity the 2D case. For a noninteracting
system the Hamiltonian is separable, $\hat H=\hat H_x+ \hat
H_y$, and hence the eigenvalues can be found independently in the two
cartesian directions,
${\cal E}_{\{i\}}={\cal E}_{i_x}+{\cal E}_{i_y}$, 
taking $\{i\}=(i_x,i_y)$ as a set of quantum numbers. Since the
wavefunction factorizes we can write the density profile as
\begin{equation}
n^{2D}_{E_F}(x,y)=n^{2D}_{I_x,I_y}=\sum_{i_x=1}^{I_x+1}\sum_{i_y=1}^{\tilde
I_y+1}|\psi_{i_x}(x)|^2|\psi_{i_y}(y)|^2 \;,
\label{dens_2D}
\end{equation}
where the upper indices $I_x$, $\tilde
I_y$ of summation are fixed by the value of the Fermi energy from the
implicit relations $E_F={\cal E}_{(I_x,1)}$ and
$E_F={\cal E}_{(i_x, \tilde I_y)}$, which we assume to be able to
solve; by definition, $\tilde I_y$ depends on the value of the quantum
number $i_x$ and its highest value $I_y$ is obtained by setting $i_x=1$.
In Eq.~(\ref{dens_2D}) we have chosen to index the
levels $(i_x,i_y)$ starting from $(1,1)$.
 
We are thus led to the final result that the 2D density profile can be
rewritten in the form of $1D$ density profile for $I_x+1$ particles in the
$x$ direction, where the $i_x$-th term is weighted by the $1D$ density profile
of $\tilde I_y +1$ particles in the $y$ direction,
\begin{equation}
n^{2D}_{I_x,I_y}(x,y)=\sum_{i_x=1}^{I_x+1} \langle\psi_{i_x}|
\delta(x-x_{i_x})|\psi_{i_x}\rangle \, n^{1D}_{\tilde I_y}(y)\;.
\end{equation}
This expression allows us to apply recursively the Green's function
method in coordinate space: we first obtain the 1D density 
profiles~\cite{noi} in
the $y$ direction,
\begin{eqnarray}
n^{1D}_{\tilde I_y}(y)=\sum_{i_y=1}^{\tilde
I_y+1} \langle\psi_{i_y}| \delta(y-y_{i_y})|\psi_{i_y}\rangle
\nonumber \\ =-\frac{1}{\pi}\lim_{\varepsilon\rightarrow 0^+}{\rm
Im}{\rm Tr}_{\tilde I_y+1}
[\hat{G}(y)]\;,
\label{dens_1D}
\end{eqnarray}
where $\hat{G}(y)=(y+i\varepsilon-\hat{y})^{-1}$ is the Green's
function and $\hat{y}$ is the position operator  in the $y$
direction expressed in the basis of the single-particle energy
eigenstates, and the trace runs over the first $\tilde I_y +1$ elements of
the matrix. We can then obtain the full 2D density profile by taking
the trace
\begin{equation}
n^{2D}_{I_x,I_y}(x,y)=
-\frac{1}{\pi}\lim_{\varepsilon\rightarrow 0^+}{\rm Im}{\rm Tr}_{I_x+1}
[\hat{G}(x){\cal N}^{1D}_{\tilde{I}_y}(y)]\;.
\label{hand}
\end{equation}
Here $\hat{G}(x)=(x+i\varepsilon-\hat{x})^{-1}$ 
is the Green's function operator in the $x$ direction, 
and ${\cal N}^{1D}_{\tilde{I}_y}(y)$ is
a diagonal ``weight'' matrix whose $i_x$-th element is given by
$n^{1D}_{\tilde I_y}(y)$. Again, the trace is taken only over the
first $I_x+1$ levels.

We have applied this general method to the case of 
2D  harmonic confinement.
The Hamiltonian is given by
\begin{equation}
\hat{H}=\hbar\omega_x(\hat{n}_x+\frac{1}{2})+
\hbar\omega_y(\hat{n}_y+\frac{1}{2}),
\label{hamilt_2D}
\end{equation}
where $\hat{n}_{x,y}=\hat{a}^\dag_{x,y} \hat{a}_{x,y}$ are the number operators
with $\hat{a}_{x,y}\,|\,\psi_{i_{x,y}}\rangle=\sqrt{i_{x,y}-1}\,|\,
\psi_{i_{x,y}-1}\rangle$ and
$\hat{a}^{\dag}_{x,y}\,|\,\psi_{i_{x,y}}\rangle=
\sqrt{i_{x,y}}\,|\,\psi_{i_{x,y}+1}\rangle$,
$\psi_{i_{x,y}}$ being respectively the wave functions of the
Hamiltonians $\hat{H}_{x,y}$ of the 
$i_{x,y}$-th energy levels, and we have considered the case
of a generic anisotropy $k=\omega_y/\omega_x$.
For this system the energy levels
${\cal E}_{i_x,i_y}$ are known and the upper limits for the sums
appearing in 
Eq.~(\ref{dens_2D}) can be evaluated explicitly:
 $I_x$  is defined by the relation
\begin{equation}
I_x={\rm int}\left[\frac{E_F}{\hbar\omega_x}-\frac{1}{2}(k+1)\right]
\end{equation}
and analogously  $\tilde I_y$  reads 
\begin{equation}
\tilde I_y={\rm int}\left[\frac{I_x+1-i_x}{k}\right].
\end{equation}

Eq.~(\ref{hand}) allows us to evaluate efficiently the 2D density
profile by exploiting the relation between the trace of a 
matrix and the determinant
of its inverse~\cite{noi}:
\begin{eqnarray}
&&
n^{2D}_{I_x,I_y}(x,y)=\nonumber\\&&
-\frac{1}{\pi}\lim_{\varepsilon\rightarrow 0^+} {\rm Im}
\frac{\partial}{\partial\lambda}\left[\det(x+i\varepsilon-\hat{x}+\lambda
\,{\cal N}^{1D}_{\tilde{I}_y}(y){\Bbb{I}}_{I_x+1})\right]_{\lambda=0}
\label{holtz}
\end{eqnarray}
where  
${\Bbb{I}}_{I_x+1}$ is a matrix with the first $I_x+1$ diagonal elements
equal to 1 and null elsewhere. 
The calculation of the 2D density profile thus reduces
to evaluating the determinant of a tridiagonal matrix, which can 
be expressed with a recursion relation as the product of infinite
terms:
\begin{equation}
\det(x+i\varepsilon-\hat{x}+\lambda\,
{\cal N}^{1D}_{\tilde{I}_y}(y){\Bbb{I}}_{I_x+1})=
\prod_{j=1}^{\infty}(x-a_j+i\varepsilon)
\label{recursion}
\end{equation}
where the factors
can be written as
$a_1=\lambda n^{1D}_{I_y}$, for $1<j\le I_x+1$,
$a_{j}=\lambda n^{1D}_{\tilde{I}_y}(y)+j/(2(x-a_{j-1}))$
and for $j>I_x+1$, $a_{j}=j/(2(x-a_{j-1}))$.
In the latter expression we have scaled the coordinate $x$ 
in units of the harmonic oscillator length $l_x$ in the $x$ direction, 
$n^{1D}(y)$ in units of 
$l_y^{-1}$ and the
resulting 2D profile in units of $(l_xl_y)^{-1}$, with $l_{x,y}=
(\hbar/m\omega_{x,y})^{1/2}$.

\paragraph*{The 3D system.}
The procedure outlined above can be applied recursively
to describe the physical case of three spatial dimensions.
In the case of noninteracting fermions the Hamiltonian
is separable in the three directions and the density 
profile can be written as
\begin{eqnarray}
&&n^{3D}_{E_F}(x,y,z)=n^{3D}_{I_x,I_y,I_z}(x,y,z)=\nonumber\\
&&
\sum_{i_z=1}^{I_z+1}\sum_{i_x=1}^{\tilde I_x+1}\sum_{i_y=1}^{\tilde{
\tilde I}_y+1}|\psi_{i_z}(z)|^2|\psi_{i_x}(x)|^2|\psi_{i_y}(y)|^2 \;,
\label{dens_3D}
\end{eqnarray}
where the highest indices of the sum are fixed by the 
implicit relations
$E_F={\cal E}_{(I_z,1,1)}$,
$E_F={\cal E}_{(i_z, \tilde I_x,1)}$ and 
$E_F={\cal E}_{(i_z, i_x,\tilde{\tilde I}_y)}$ and $I_x$ and $I_y$
are the highest value of $\tilde I_x$ and $\tilde{\tilde I}_y$. 
Analogously to the case of the 2D confinement, 
the 3D particle density (\ref{dens_3D})
can be reduced to an expression for an
effective 1D density profile for $I_z+1$ particles, 
with the $i_z$-th term weighted by the 2D density profile 
$n^{2D}_{\tilde I_x,\tilde I_y}(x,y)$ 
\begin{eqnarray}
n^{3D}_{I_z,I_x,I_y}(x,y,z)&=&
\sum_{i_z=1}^{I_z+1} \langle\psi_{i_z}|
\delta(z-z_{i_z})|\psi_{i_z}\rangle \, n^{2D}_{\tilde I_x,\tilde I_y}(x,y)
\nonumber\\
&=&-\frac{1}{\pi}\lim_{\varepsilon\rightarrow 0^+}{\rm Im}{\rm Tr}_{I_z+1}
[\hat{G}(z){\cal N}^{2D}_{\tilde{I}_x,\tilde{I}_y}(x,y)]\; .
\label{dens_3D_2}
\end{eqnarray}
Here, $\hat{G}(z)$ is the Green's function for the $\hat z$ operator,
${\cal N}^{2D}_{\tilde{I}_x,\tilde{I}_y}(x,y)$ is a diagonal
matrix whose $i_z$-th element is given by 
$n^{2D}_{\tilde{I}_x,\tilde{I}_y}(x,y)$ and 
$\tilde I_y=\tilde{\tilde I}_y(i_x=1;i_z)$ is the
highest value of $\tilde{\tilde I}_y$ at fixed $i_z$.

In the case of harmonic confinement, the indices of the occupied levels 
are functions of the anisotropy parameters $k=\omega_y/\omega_x$
and $l=\omega_x/\omega_z$ of the trap,
\begin{equation*}
I_z={\rm int}\left[\frac{E_F}{\hbar\omega_z}-\frac{1}{2}(lk+k+1)\right],
\end{equation*}
\begin{equation*}
\tilde I_x
={\rm int}\left[\frac{I_z+1-i_z}{l}\right],
\;\;\;{\rm and}\;\;\;
\tilde I_y
={\rm int}\left[\frac{I_z+1-i_z}{kl}\right]\,,
\end{equation*}
and the density profile, in units of $(l_xl_yl_z)^{-1}$
takes the simple form
\begin{eqnarray}
&&n^{3D}_{I_z,I_x,I_y}(z,x,y)=\nonumber\\
&&-\frac{1}{\pi}\lim_{\varepsilon\rightarrow 0^+}{\rm Im}
\frac{\partial}{\partial\lambda}\left[\prod_{j=1}^\infty(z-a_j
+i\varepsilon)\right]_{\lambda=0}\,,
\label{rec3}
\end{eqnarray}
where  the effect of the presence of the confinement in the
$\{x,y\}$ plane is taken into account by the $a_j$ factors, with
$a_1=\lambda n^{2D}_{I_x,I_y}(x,y)$, for $1<j\le I_z+1$,
$a_{j}=\lambda n^{2D}_{\tilde{I}_x,\tilde{I}_y}(x,y)+j/(2(z-a_{j-1}))$
and for $j>I_z+1$, $a_{j}=j/(2(z-a_{j-1}))$.
In Eq.~(\ref{rec3}) the coordinates are in units of the harmonic oscillator
lengths $l_x$, $l_y$ and $l_z$.

\paragraph*{Numerical results.}
We show here some results for the density profiles
in the case of a 2D harmonic trap for various values of the
anisotropy parameter.
Our 2D results correspond to column densities of 3D systems
in which only one level is occupied in the integrated  $z$ direction.
Our numerical procedure is the following.
The first step is to calculate and store the 1D density profiles 
from 1 to $\tilde{I}_y+1$ particles.
Then, the scheme given in Eqs.~(\ref{holtz})-(\ref{recursion}) is 
easily implemented,
performing the calculation of the determinant~(\ref{recursion}) up to the 
product of $M$ terms. This approximation, which corresponds to neglect
all the excited states larger than $M$, is the same 
as the one used for the evaluation
of the diagonal elements of the matrix ${\cal N}^{1D}_{\tilde{I}_y}(y)$.
Our calculations have been performed using $M=10^6$ and $\varepsilon=0.01$.

\begin{figure}
\centering{
\epsfig{file=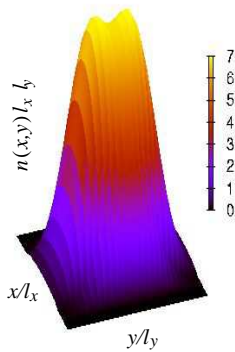,width=0.5\linewidth}}
\caption{Density profile of 1050 fermions in a 2D harmonic trap
as function of $x/l_x$ and $y/l_y$, with 
$E_F=875\,\hbar\omega_x$ and $k=350$.}
\label{fig1050}
\end{figure}

\begin{figure}
\centering{
\epsfig{file=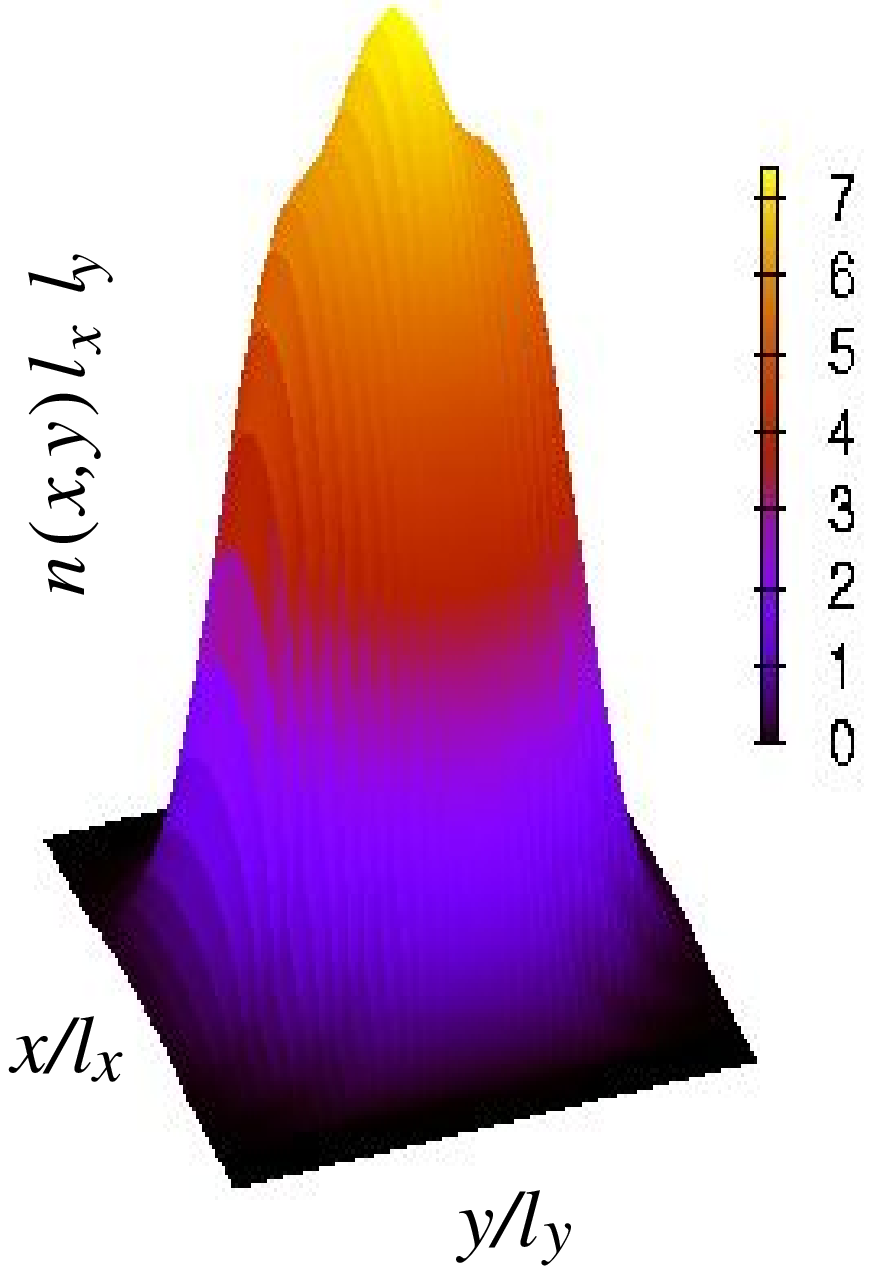,width=0.5\linewidth}}
\caption{Density profile of 1038 fermions in a 2D harmonic trap
as function of $x/l_x$ and $y/l_y$, with 
$E_F=605\,\hbar\omega_x$ and $k=173$.}
\label{fig1038}
\end{figure}

\begin{figure}
\centering{
\epsfig{file=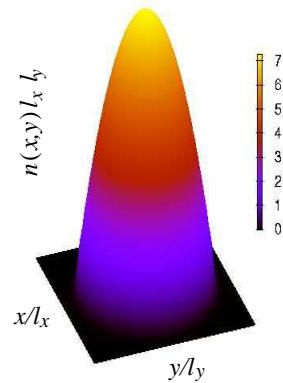,width=0.5\linewidth}}
\caption{Density profile of 1035 fermions in a 2D isotropic
harmonic trap as function of $x/l_x$ and $y/l_y$, 
with $E_F=45\,\hbar\omega_x$.}
\label{fig1035}
\end{figure}

\begin{figure}
\epsfig{file=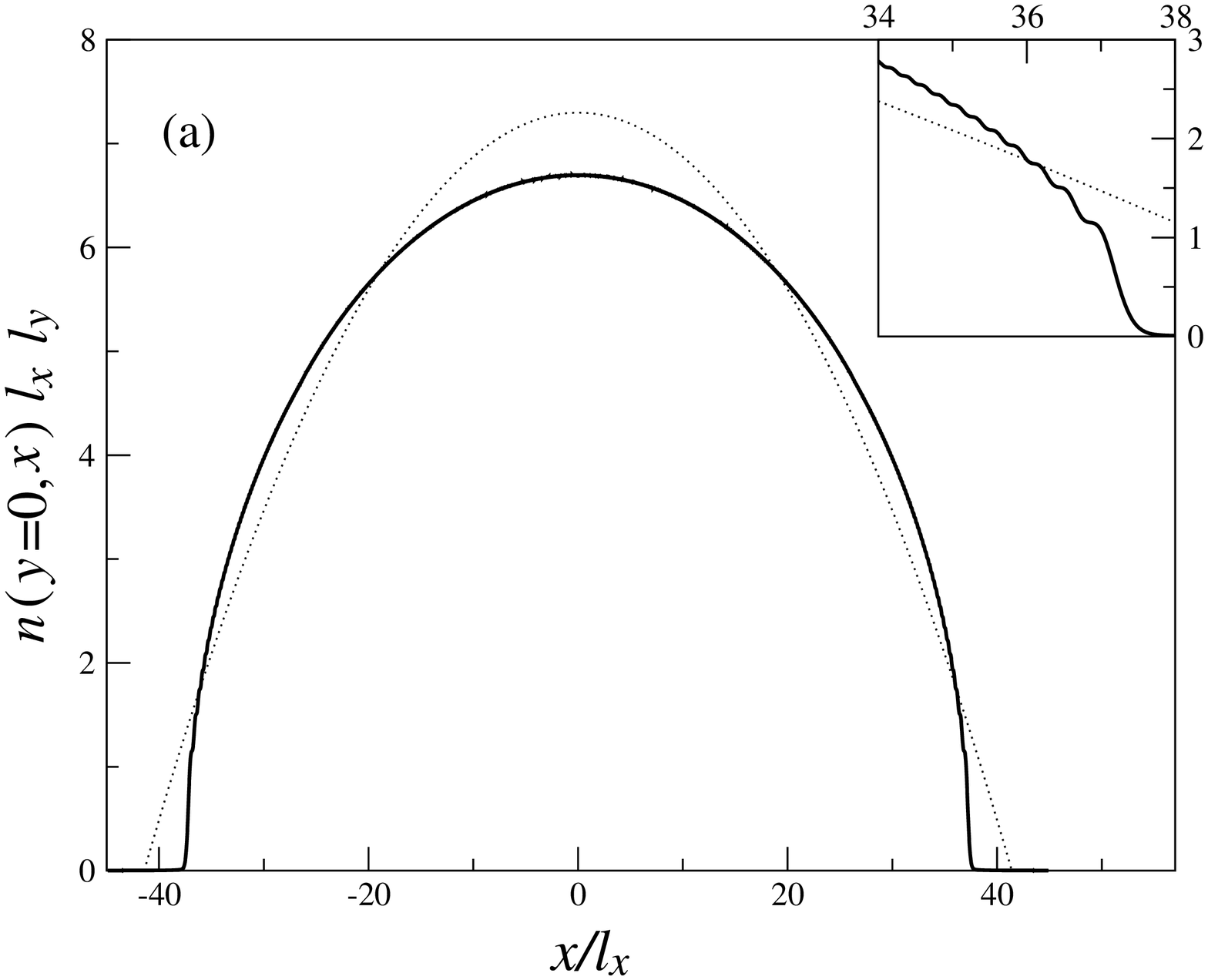,width=0.5\linewidth}
\hspace{-1cm}
\epsfig{file=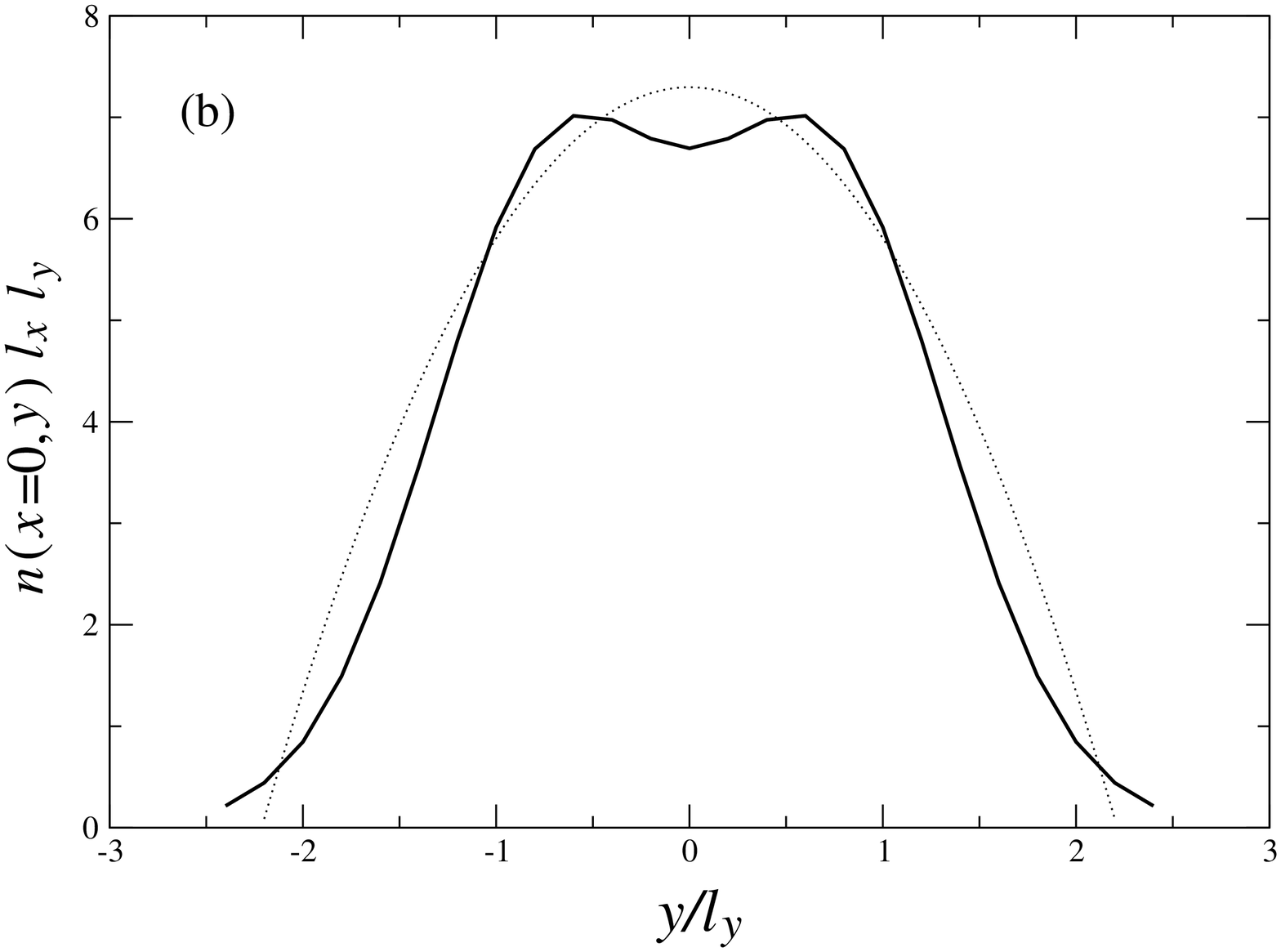,width=0.5\linewidth}
\caption{Sections of the density profile of Fig.~\ref{fig1050}
for $y=0$ (a) and for $x=0$ (b). The exact profiles (continuous line) 
are compared
with those evaluated in the Thomas-Fermi approximation (dotted line).
The inset of (a) shows a zoom of the tail.}
\label{sec_fig1050}
\end{figure}

In Figs.~\ref{fig1050}-\ref{fig1035} we report the density profiles
of three different closed-shell systems with a thousand of particles. 
Figs.~\ref{fig1050} and ~\ref{fig1038} refer to the case 
a trap with a large anisotropy, having 
only two and three levels occupied in the $y$ direction,
respectively. The prominent shell structure which comes out in the direction
of tight confinement is completely lost in the case of an isotropic
trap for such a large number of fermions (Fig.~\ref{fig1035}).
Another peculiarity of the system with a large anisotropy is that the tails
of the profiles in the direction of weak confinement have a 1D
behavior: for large values of $x$ only the Hermite polynomials of
high degree have a significant weight, 
but these terms (the last $I_x+1-kI_y$ ones) are weighted by the same factor
- the density of a single fermion in the $y$ direction - giving rise to
an essentially 1D profile in the tails
as is shown in the 
inset of Fig.~\ref{sec_fig1050}(a). 
However, in the case of a mesoscopic
system where a large number of levels are occupied in the longitudinal 
direction,
the 1D shell structure is not well visible.

We have also compared the exact profiles with those given in the 
Thomas-Fermi (TF)
approximation for $N$ fermions at anisotropy parameter $k$
\begin{eqnarray}
n^{2D}_{TF}(x,y)&=&
\frac{m}{2\pi\hbar^2}(E_F-m\omega_x^2(x^2+k^2y^2)/2)\nonumber\\
&\times&\theta(E_F-m\omega_x^2(x^2+k^2y^2)/2)
\end{eqnarray}
with $E_F=(2kN)^{1/2}\hbar\omega_x$ evaluated in the TF limit.
While for the symmetric case ($k=1$) and $N\simeq 1000$
the TF profile is practically 
indistinguishable from the exact one, 
the local density approach
completely fails in describing not only the narrow profile in the $y$ direction
(see Fig.~\ref{sec_fig1050}(b)) but also the profile in the $x$ direction 
(Fig.~\ref{sec_fig1050}(a)) 
even for a large number of occupied levels. 

In conclusion in this Letter we have given a general formula for the exact
particle density of 2D and 3D fermions in external confinement
in terms of a Green's function in coordinate space.  
Our approach allows us to treat systems with arbitrary
anisotropy since we can deal separately with each
cartesian direction by reducing recursively the problem
to the evaluation of onedimensional profiles, where the
other degrees of freedom are taken into account through
renormalization factors.
The same idea can be exploited to evaluate 
the higher-order moments of the density matrix.
We have used this method, which is particularly suited to the case of harmonic 
confinement, to evaluate 2D density profiles of mesoscopic systems
for various values of the anisotropy parameter.
In the case of large anisotropy we obtain a prominent shell structure,
which should be experimentally observable in mesoscopic 
clouds at temperatures 
lower than the harmonic-oscillator energy spacing in the tight direction
\cite{noiT}.
For such anisotropic systems we have found that the 
local density approximation
completely fails in reproducing the exact profiles even for a large
number of atoms ($N\simeq 1000$). 
As a consequence, we expect that the TF density functional,
which has been shown to work well for isotropic traps~\cite{brack} 
should be inadequate to deduce the kinetic energy density in the case
of large anisotropy.

\acknowledgements
We are indebted to Prof. M.P. Tosi for many useful discussions and
encouragement.
We acknowledge support from INFM through the PRA2001 program.

\end{document}